\documentclass[letterpaper, 10 pt, conference]{ieeeconf}  

\IEEEoverridecommandlockouts                              

\usepackage{amsmath,graphicx,xcolor,siunitx}
\usepackage[acronym,nonumberlist,nopostdot,nogroupskip]{glossaries}
\loadglsentries{glossary.sty}

\title{\LARGE \bf
Systematic Hardware Integration Testing for Smart Video-based Medical Device Prototypes
}

\author{Oliver Bause$^{1}$, Julia Werner$^{1}$, and Oliver Bringmann$^{1}$
\thanks{This work has been partly funded by the German Federal Ministry of Education and Research (BMBF) in the project MEDGE (16ME0530).}
\thanks{$^{1}$Department of Computer Science, University of Tübingen, Tübingen, Germany, {\tt\small oliver.bause@uni-tuebingen.de}}%
}

\begin{document}

\maketitle
\thispagestyle{empty}
\pagestyle{empty}

\begin{abstract}
This paper presents a \gls{hil} verification system for intelligent, camera-based in-body medical devices.
A case study of a \gls{vce} prototype is used to illustrate the system's functionality.
The \gls{fpga}-based approach simulates the capsule's traversal through the \gls{gi} tract by injecting on-demand pre-recorded images from \gls{vce} studies.
It is demonstrated that the \gls{hil} configuration is capable of meeting the real-time requirements of the prototypes and automatically identifying errors.
The integration of \gls{ml} hardware accelerators within medical devices can be facilitated by utilising this configuration, as it enables the verification of its functionality prior to the initiation of clinical testing.
\end{abstract}

\section{Introduction}

Medical devices used within the human body play a crucial role in modern healthcare, particularly in diagnosis and treatment.
Implantable devices, such as pacemakers or cochlear implants, are designed to restore or support essential bodily functions over the long term \cite{joung2013development}.
While \gls{vce} represents a procedure in the field of \gls{gi} diagnostics, offering a minimal invasive method for visualizing the entire \gls{gi} tract.
Specifically, the examination of the small intestine remains difficult, since this region is traditionally only hardly accessible with conventional endoscopy techniques.
First introduced in 2000 by Iddan et al. \cite{iddan2000wireless}, \gls{vce} has transformed the approach of diagnosing small bowel disorders, particularly in cases of obscure gastrointestinal bleeding \cite{patel2022obscure}, Crohn's disease \cite{goran2018capsule}, and small bowel tumors \cite{cheung2016usefulness}.
It involves swallowing a small, pill-sized camera that naturally traverses the \gls{gi} tract.
The captured images are transmitted to an external receiver for a following detailed expert analysis.

Most \gls{vce} devices are equipped with a small wide-angle camera, an LED for illumination, a \gls{mcu} or an \gls{fpga}, and a transmitter.
The operation of these systems is characterized by a frame rate ranging from one to six \gls{fps}, with a resolution less than 0.3 \gls{mp} \cite{chetcuti2021capsule}.
Some video capsules used for upper \gls{gi} endoscopy have a temporary frame rate of up to 35 \gls{fps}, though limiting their battery life to less than 90 minutes.
The restricted volume within the capsule has a significant impact on the battery's capacity.
Consequently, the battery life of these capsules is typically limited to twelve hours.
Nevertheless, in certain instances, the capsule may require in excess of a day to traverse the \gls{gi} tract.
As this might come at the cost of an incomplete screening procedure, improvements for the overall \gls{vce} procedures are still needed.

Recent advancements in \gls{vce} technology, including improvements in image resolution, battery life, and data processing algorithms, have further strengthened their diagnostic capabilities \cite{chetcuti2021capsule}.
These innovations have expanded the clinical applications of \gls{vce} beyond the small intestine, with newer devices designed to examine the esophagus, colon, and the entire \gls{gi} tract.
Furthermore, research is ongoing in the application of \gls{ml} for capsule endoscopy, with the potential to facilitate expert analysis \cite{chetcuti2021capsule}.
As a case in point, these tools have the potential to reduce the time required for the detection of lesions in an entire small bowel video from up to 90 minutes to less than 30 minutes \cite{tziortziotis2021role, gilabert2022artificial}.
However, it should be noted that these approaches invariably employ \gls{ml} for the purpose of evaluating the video output of the \gls{vce} subsequent to its traversal of the \gls{gi} tract.
Integrating \gls{ml} models into the capsule's \gls{isp} by deploying a custom low-power hardware accelerator can be used to perform on-board image classification, which allows to directly decide whether an image should be transmitted for further inspection.
For example, Werner et al. \cite{werner2023precise} presented a small model to determine the current position of the capsule within the \gls{gi} tract on-site.
Execution of such models on the device directly facilitates the more selective utilisation of the capsule's constrained battery capacity for the purpose of selectively capturing \gls{gi} tract segments of relevance.
Additionally, integrating anomaly detection might be feasible, enabling the temporary increase of the capsule's frame rate to collect more images of potential anomalies.
These could then be flagged and, during diagnosis, multiple frames of the same scene could be used to reconstruct a single image with a higher resolution \cite{farsiu2005multiframe}.

The deployment of \gls{ml} into medical devices, such as video capsules, presents a series of significant challenges.
Due to the high cost associated with the fabrication of a prototype, a meticulous evaluation and verification of the software hardware co-design's functionality is imperative prior to proceeding.
The utilization of a defective \gls{ml} model or an unoptimized design can potentially lead to a reduction in the quality of the captured video. 
In a worst-case scenario, this may result in the failure to record the relevant segments of the \gls{gi} tract entirely.
Therefore, a \gls{hil} test setup is required, that can emulate the sensor input and evaluate the capsules output in real-time.
Pre-recorded videos from existing capsules, an artificial simulation of the \gls{gi} tract, or a mixture of both methods could be used to generate the emulated camera feed.
This paper proposes a \gls{fpga}-based \gls{hil}, that verifies the operational integrity of the device's \gls{isp} pipeline within the context of employing a \gls{ml} hardware accelerator.

\section{Testing of In-Body Medical Devices}

Preclinical development and testing of new medical devices is costly and time consuming.
Multiple "design-build-test-redesign" cycles need to be performed before reaching the initial clinical testing, which is a major step towards the regulatory approval of a medical device \cite{kaplanMedicalDeviceDevelopment2004}.
Before the initial trial, patients cannot be involved into the prototype testing.
Consequently, animal testing is frequently employed as an alternative.
However, integrating a \gls{hil} into this process can reduce the number of iteration cycles by simulating scenarios from real patients.

\begin{figure}[t]
  \centering
  \centerline{\includegraphics[width=8.5cm]{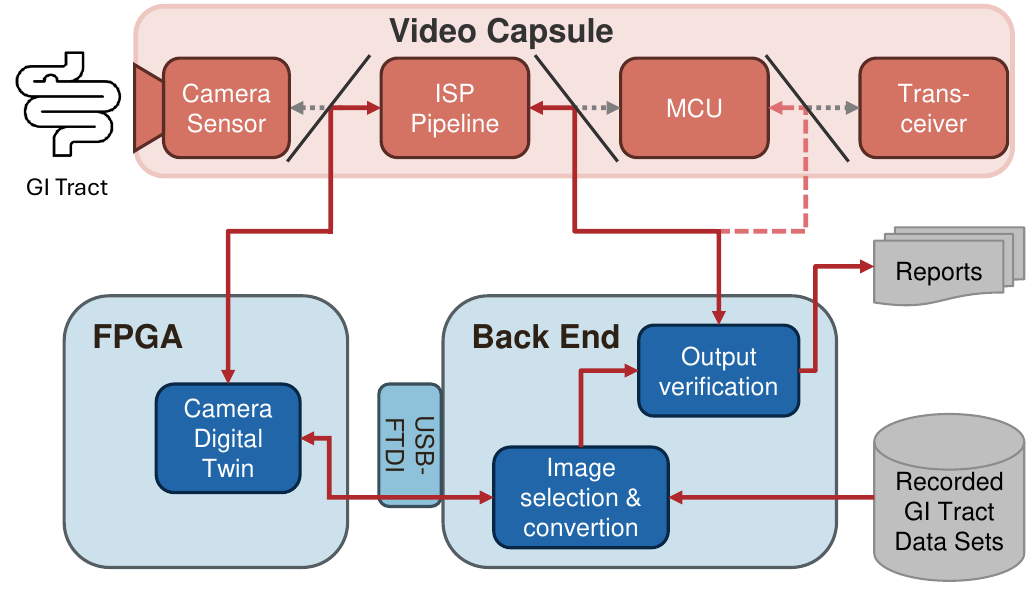}}
\caption{Top level diagram of a VCE device and the proposed HIL framework.}
\label{fig:framework}
\end{figure}

The \gls{dut} of our presented \gls{hil}, a \gls{vce} prototype, will feature two kind of interfaces to interact with its environment.
First, it is equipped with an array of different sensors.
A video capsule can be provided with one to up to four cameras for an one- or bi-directional coverage \cite{chetcuti2021capsule}.
These miniature cameras capture images with a Bayer color filter array.
Thus, demosaicing needs to be performed to obtain the RGB image.
In order to reduce the energy, memory, and chip area requirements within the capsule, the RAW images of the \gls{gi} tract are transmitted to an \gls{obd}.
Finally, the RGB images are reconstructed for the following medical expert analysis.
Additionally, miscellaneous sensor configurations, like pressure, temperature, or pH sensors, could theoretically be added to further contribute metadata to the captured images.
This can either be used to assist the analysis or enhance the control decisions of the capsule itself.
The PillCam SB3, for example, offers a variable frame rate from two to six \gls{fps} depending on the current velocity of the capsule \cite{zwingerCapsoCamSV1PillCam2019}.
As illustrated in Figure \ref{fig:framework}, the \gls{hil} needs to emulate the sensor input in real time by deploying a digital twin of the aforementioned sensor.
It can either replay a previously recorded data set or utilize a virtual simulation environment.
It could also be possible for a combination of both to be employed.
If the \gls{dut} is equipped with different senor types, it is crucial to obtain a data set that also has meaningful data points for each senor.

The second interface is constituted by wireless communication with the \gls{obd} transceiver, given that only a small number of capsules save images on an on-board data storage system \cite{friedrichFirstClinicalTrial2013}.
Thus, these capsules with their data need to be retrieved after leaving the body.
However, most \gls{vce} devices require the patient to wear an \gls{obd} as the images are constantly transmitted by the capsule.
Some receiver analyze the incoming video stream to, for example, adjust the configuration of the capsule or instruct the patient to consume more laxative boosters \cite{chetcuti2021capsule}.
Consequently, the \gls{hil} must additionally intercept the output of either the entire capsule or its submodules, as demonstrated in Figure \ref{fig:framework}. 
A comparison of the output with the previously injected input can then be performed.

A major design challenge while developing \gls{vce} devices is the limited, available volume within the capsule that needs to fit the sensors, the control unit, the transceiver, and a battery pack powerful enough for the capsule to last at least twelve hours.
Therefore, our \gls{dut} will include an ultra low-power \gls{ml} hardware accelerator.
The device should perform on-the-edge image classification to evaluate the location and possible points of interest within the \gls{gi} tract.
The objective is to leverage the limited energy in an efficient and purposeful manner.
Hence, particular emphasis should be placed on the validation of the hardware functionality and the integrity of the implemented model.
As the video capsule is used within the human body and is single-use only, it must be assured that the \gls{ml} model does not result in potential data loss.

\subsection{Related Work}
To the best of our current knowledge, there are no other \gls{hil} publications for camera-based medical devices.
Even exhaustive surveys of the current developments of \gls{vce}s \cite{chetcuti2021capsule,goran2018capsule} do not cover the testing and functional verification of the capsules.
However, Mosleh at al. \cite{mosleh2020hardware} developed a \gls{hil} for \gls{isp} pipelines to optimize their hyperparameters for object detection and segmentation tasks.
They use a given \gls{isp} pipeline, feed it with raw image data, and calculate the loss to determine the quality of the selected hyperparameters.
Additionally, Komorkiewicz et al. \cite{komorkiewiczFPGAbasedHardwareintheLoopEnvironment2016} also published an automotive \gls{hil} setup that uses an \gls{fpga} to inject a video stream into an onboard processing unit.
Since their setup is optimized for the requirements of testing advanced driver assistance systems with commercial off-the-shelf hardware components, it can not be used to verify small in-body medical devices.
Also, domain specific segments of the \gls{hil} like the CAN frame injection are not applicable in our use-case.

Mascio et al. \cite{mascioHardwareLoopImplementation2020} published a \gls{hil} for medical devices interacting with the human heart.
They developed several extensive oscillator-based models to simulate different heart conditions.
However, like \cite{mosleh2020hardware}, they did not cover the actual hardware implementation of their presented \gls{hil}.
By analyzing their provided graphics, we can only speculate, that they utilized a Raspberry Pi minicomputer to facilitate the control of the oscillator.

Our methodology diverges from the aforementioned approaches as our objective is to optimize the hardware-software co-design of the \gls{isp} pipeline with an integrated \gls{ml} hardware accelerator to be used within an in-body medical device.
Consequently, our aim is to develop a \gls{hil} that not only facilitates the appropriate selection of hyperparameters but also enables the optimization of the \gls{isp} pipeline's architectural design.
Simultaneously, the presented \gls{hil} focuses on the unique challenges and requirements of medical devices with real-time image injection.
Unlike the automotive setting, where early testing can be performed with a non-intrusive co-pilot system within an existing vehicle, the medical environment requires many regulations \cite{RegulationEU20172017} before deploying clinical trials with human subjects.

\section{Hardware-in-the-Loop Framework}

The presented \gls{hil} framework is composed of two main components, the back-end, which is executed on a standard Linux computer and the \gls{fpga}-based sensor emulation. 
Our database consists of multiple \gls{vce} studies from the Rhode Island \cite{charoenRhodeIslandGastroenterology2022a} and the Galar data set \cite{leflochGalarLargeMultilabel2024}.
Each study has multiple thousand chronologically sorted RGB images, that are labeled according to their location within the \gls{gi} tract.
The digital twin of the \gls{dut}'s camera sensor, an AMS NanEyeC miniature camera module \cite{naneyec}, needs to send the correct image to the \gls{dut} with respect to the current simulation time stamp and the possibly variable frame rate of the capsule.
As integrated memory space is tightly constrained on \glspl{fpga}, it is not realizable to store the whole data set of a single study on the \gls{fpga} itself.
The variable frame rate of the capsule's camera module prevents a calculated pre-loading of images, thus, an on-demand fetching is implemented to simulate the camera input.

\subsection{FPGA Design}
\label{sec:fpga}

Different hardware platforms to inject the image stream were evaluated.
The camera operates in slave mode with an external clock source for the communication \cite{naneyec}.
Therefore, the virtual twin must also react to an external controller.
The majority of microcontrollers support only an efficient but inflexible hardware implementation of the master side of common serial communication protocols.
Thus, a \gls{gpio}-based software implementation would be needed to realize the digital twin of the camera.
However, the software overhead limits the maximum available clock frequency to less than \SI{2}{\MHz} \cite{softwarespi} which is much lower than the NanEyeC's maximum clock frequency of \SI{75}{\MHz}.
The FTDI USB-to-SPI slave module had similar limitations.
The communication protocol could not be defined dynamically.
As a result, it was not possible to implement the state machine-based protocol of the NanEyeC camera \cite{naneyec}.
Additionally, the USB interface cannot guarantee the bandwidth and latency which are the major requirements of the digital twin.
An \gls{fpga}, however, offers the required flexibility and can achieve the latency and real-time requirements even at a high master clock speed.

The Digilent ZYBO \gls{fpga} \cite{zybo2027} acts as digital twin of the emulated NanEyeC image sensor.
It mirrors the cameras functionality and behavior.
As a result, it can be connected to the \gls{vce} prototype instead of the actual camera without any system adjustments.
As illustrated in Table \ref{tab:fpga_util}, the low hardware utilisation of the presented \gls{hil} facilitates the possible integration of further digital twins of sensors.
The model has been written in SystemVerilog and designed modularly to allow an extension to be straightforward.
Moreover, it is feasible to incorporate the module into the electronic circuit simulation of the \gls{vce} design itself prior to the prototype manufacturing stage.

\begin{table}[t]
\caption{Resource utilization of flip-flops (FF), lookup tables (LUT), slices, and block RAM (BRAM) of the ZYBO FPGA.}
\label{tab:fpga_util}
\begin{center}
\begin{tabular}{|c|c|c|c|c|}
\hline
    & \textbf{FF} & \textbf{LUT} & \textbf{SLICE} & \textbf{BRAM} \\
\hline
Used & 209 & 461 & 168 & 32 \\
\hline
Available & 35,200 & 17,600 & 4,400 & 60 \\
\hline
Utilization & 1\% & 3\% & 4\% & 53\% \\
\hline
\end{tabular}
\end{center}
%
%
\end{table}

The NanEyeC camera sensor utilizes a serial dual-wire communication protocol with one clock and one bi-directional data line.
The camera offers a wide range of exposure settings, can be set to idle and a reset is performed by executing a power cycle.
After leaving the idle mode and each power-on, the first frame should be discarded.
The camera has a rolling shutter, so pixel that are not in reset will be otherwise overexposed.

\begin{figure}[t]
\centering
\centerline{\includegraphics[width=8.5cm]{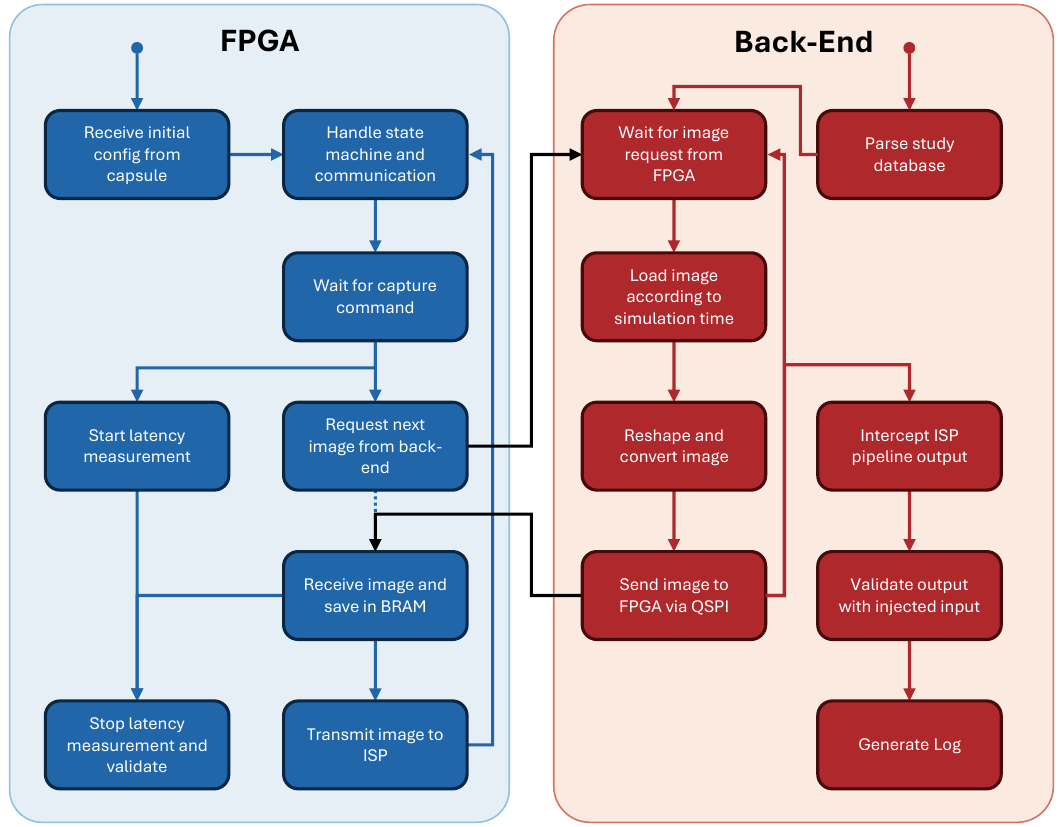}}
\caption{State machine of the task of the \gls{fpga} and the back-end and the interaction between them.}
\label{fig:fsm}
\end{figure}

The camera has a 320 by 320 pixel resolution with a single 10-bit channel in a Bayer pattern.
It can be configured with up to 58 frames per second at a clock frequency of \SI{75}{\MHz}, resulting in a maximum image data transfer of \SI{59.392}{Mb/s}.
However, \gls{vce} devices mostly operate in a single digit frame rate.
So, the real-world application requirements are even lower at less than \SI{2}{Mb/s}.
The \gls{fpga} can only store one image at a time, as the RAM utilization is already at \SI{53}{\%} (see Table \ref{tab:fpga_util}).
A selected number of \gls{fpga} boards offer onboard DDR memory.
However, the application of this solution is also limited due to the fact that the compressed data set for a single patient has been known to exceed multiple gigabytes, thus also exceeding the capacity of the DDR memory.
Therefore, a critical requirement that must be met by the aforementioned on-demand fetching approach is the response time of the camera, which is dependent on the clock speed and the configured exposure settings.
With the lowest exposure settings, the image transmission of the NanEyeC's internal communication state machine begins after \SI{11520} clock cycles after leaving the idle mode \cite{naneyec}.
Nonetheless, it is recommended that the initial image following each idle period is disregarded due to the rolling shutter artefact.
In order to conserve energy, the rolling shutter does not operate during idle mode.
Consequently, in order to achieve accurate exposure times, it is necessary for each pixel row to be reset. Therefore, the actual number of clock cycles preceding the first usable image transmission is \SI{1298880}, as the camera's state machine needs to complete a whole cycle of its state machine to properly reset all pixels.
At a clock frequency of \SI{75}{\MHz}, the resulting time frame is \SI{17}{\ms}.
As illustrated in Figure \ref{fig:fsm}, within this time frame, the \gls{fpga} needs to retrieve the image from the database that is connected to the back-end computer.
The digital twin also measures the response time of the back-end to ensure that the requested image arrived in time.
Otherwise, a flag will be raised to notify the back end of the failed image transmission.
Consequently, the output of the image pipeline will be false which needs to be accounted by the evaluation of the back end.
Additionally, the \gls{fpga} also logs the time spent in idle and active mode.
This enables a preliminary power estimation of the selected camera settings after the simulation is completed.

\subsection{Back-End}
The back-end provides the images to the \gls{fpga} and verifies the output of the \gls{vce} prototype and its submodules.
It is connected to an online database that has multiple \gls{vce} studies \cite{charoenRhodeIslandGastroenterology2022a, leflochGalarLargeMultilabel2024}.
The images of these data sets are stored either as PNG or JPEG files.
Depending on the used capsule that generated these data sets, the images need first to be resized to the NanEyeC's resolution of 320 by 320 pixels.
In the data sets, the images have either the same or a higher resolution that needs to be compressed.
After that, the loaded image is converted back into the Bayer pattern.
A considerable number of VCE devices are equipped with diminutive image sensors that employ the identical RGGB Bayer pattern.
The receiver or the software that displays the video to the medical experts converts them into a conventional RGB image that is interpretable by humans.
It is evident that, given the framework's exclusive extraction of the original sensor data, accompanied by the elimination of the interpolated colour channel, there is negligible to no loss of original sensor data.

The \gls{hil} also enables an automatic verification of the capsule's image pipeline.
Medical use cases have high functional safety requirements.
To validate the neural network that is executed on the integrated \gls{ml} hardware accelerator, the output can be compared to the emulated input.
Since the data sets are labeled, a comparison between network prediction and actual label of the given input is straight forward.
This verification step could be performed between different submodules of the capsule.
Currently, the \gls{hil} focuses on the testing of the \gls{isp} pipeline.
However, it can be extended to support the wireless receiving of the capsules output to emulate the on-body receiver or intercept the communication between the \gls{mcu} and the transceiver.

\subsection{PC to FPGA Image Transmission}

An FT4222 USB-to-SPI interface \cite{ftdi4222} is used to handle the communication between the back-end and the \gls{fpga}.
Latency being a more significant aspect than bandwidth, Quad-SPI was selected as the communication protocol.
The protocol's simplicity reduces the hardware overhead required on the \gls{fpga} in comparison to alternative protocols such as HDMI or Ethernet.
It also facilitates more straightforward extension with additional virtual sensors.

When a new image should be captured, the digital twin on the \gls{fpga} triggers an interrupt via the \gls{gpio} ports of the FT4222 device.
Thereafter, the back-end determines the subsequent image, taking into account the current simulation time stamp.
It loads and converts the selected frame with a BGGR Bayer filter. 
Converting the image on the PC reduces the amount of data transmitted by two thirds as only one channel needs to be transmitted.
It also prevents the necessity of an image convertion pipeline on the \gls{fpga}.
The image is then sent through the FT4222 SPI-Master module at \SI{40}{\MHz}.
The \gls{hil} support standard, Dual-, and even Quad-SPI.
At the maximum clock frequency of \SI{75}{\MHz} of the NanEyeC camera, Quad-SPI needs to be used to achieve the required throughput and latency.

\section{Evaluation}

The evaluation of the presented \gls{hil} is performed by integrating an early prototype of our \gls{isp} pipeline into a customized Pulpissimo \gls{soc} \cite{PULPFAQs}. 
The \gls{soc} is programmed onto a NEXYS Artix-7 A100T \gls{fpga} \cite{NexysA7Digilent} and acts as the capsule's \gls{mcu}.
The software executed on the Pulpissimo triggers a capture command of the image pipeline and forwards the received image to the back-end of the \gls{hil}.
To verify the functional correctness of the digital camera twin, the back-end compares the received output image pixel-wise with the originally injected one.
Differences will be logged to identify possible faults within the \gls{hil} or the \gls{isp} pipeline under test.

\subsection{Capture Latency}

\begin{figure}[t]
  \centering
  \centerline{\includegraphics[width=8.5cm]{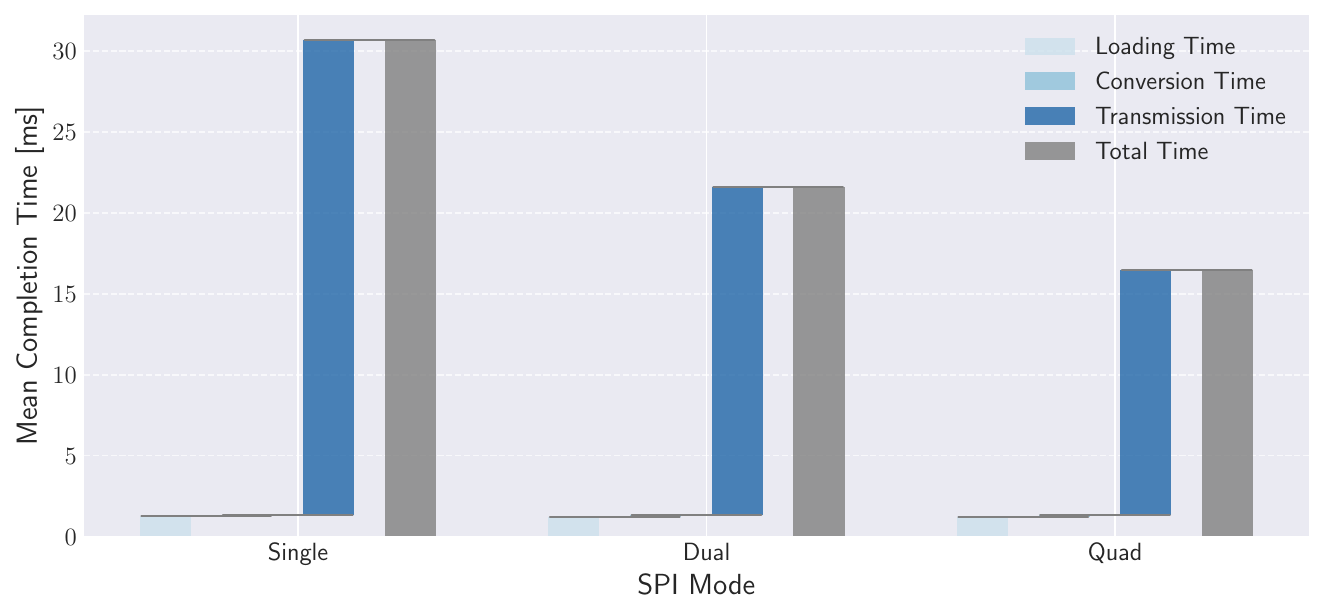}}
\caption{Comparison of the average transmission times during a simulation of a complete study with more than \SI{6300} images with Single-, Dual-, and Quad-SPI.} 
\label{fig:spi}
\end{figure}

The back-end tracks the latencies of time critical sub-tasks during the image transmission.
As shown in Figure \ref{fig:spi}, the image loading from the database and the RGB to Bayer conversion require \SI{1.22}{\ms} and \SI{0.09}{\ms} in all three SPI modes.
The transmission through the FTDI interface accounts for the majority of the runtime, with the times recorded as \SI{29.38}{\ms}, \SI{20.28}{\ms} and \SI{15.18}{\ms} for Single-, Dual- and Quad-SPI respectively.
The linear decrease in runtime suggests that a delay of approximately \SI{10}{\ms} is attributable to the overhead of the USB-to-SPI adapter's API.
Since the back-end splits the image transmission into two separate API calls, each call causes an overhead of \SI{5}{\ms}.

As described in Section \ref{sec:fpga}, if the NanEyeC camera is clocked at its maximum of \SI{75}{\MHz}, the transmission of the first pixel starts \SI{17}{\ms} after the capture command is received.
The \gls{hil} is able to send the first pixel data within \SI{6.5}{\ms} and stores the complete image in less than \SI{16.5}{\ms} on the \gls{fpga}.
Thus, the real-time response time required by the ISP to successfully inject an image stream is met.

\subsection{Fault Injection}
The digital twin has been designed to measure the time interval between transmitting the interrupt signal to the back-end and receiving the initial pixel data, the latency requirement of which is set to \SI{120}{\ms} by operating the camera module at only \SI{5}{\MHz}.
In order to subject the simulation integrity to a stress test, the response time of the back-end database was set at random to a maximum of \SI{2.5} seconds.
The utilisation of an online database can be susceptible to elevated access latency, particularly in scenarios where concurrent access is observed by other clients.

\begin{figure}[t]
  \centering
  \centerline{\includegraphics[width=8.5cm]{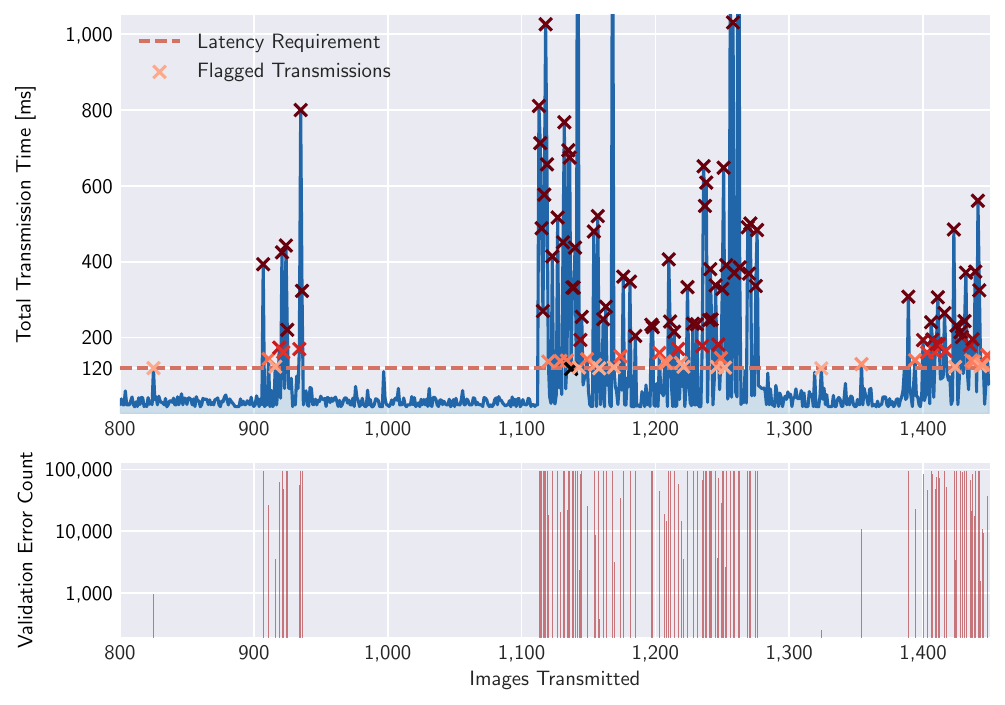}}
\caption{Identification of excessive transmission periods that lead to corrupted images being received with their respective number of pixel deviations.}
\label{fig:error}
\end{figure}

Transmissions that potentially contravene the real-time latency requirements of the camera must be identified and flagged so that the back-end can take them into account during validation of the received output.
As demonstrated in Figure \ref{fig:error}, with the exception of image \SI{1137}, which only marginally exceeded the stipulated requirement, all flagged images resulted in the transmission of a corrupted image.
Transmissions that only slightly exceeded the \SI{120}{\ms} latency exhibited a reduced pixel deviation, a phenomenon attributable to the faster bandwidth of the \gls{hil} in comparison to the bandwidth of the evaluated \gls{isp}. 
The back-end was able to catch up, and the subsequent transmitted pixels were found to be accurate. 
It is important to note that, depending on the frequency of errors, a simulation restart may be a necessary consideration.

\subsection{Power Estimation}
The power consumption of each component of the video capsule is crucial, as the battery capacity is strictly constrained. 
To calculate a preliminary power estimation of the NanEyeC module, the digital twin monitors its idle and active times. 
In conjunction with the power characteristics of the camera in both states, as outlined in the manual \cite{naneyec}, an approximate power consumption can be calculated. 
As demonstrated in Figure 5, the power consumption of the camera below one fps is less than \SI{5}{\mW}. 
Nevertheless, it increases drastically with increasing frame rates as the camera's idle times become negligible.
The stated power characteristics in the manual are at a higher clock rate, thus this estimation is more pessimistic than the values observed in real-world conditions. 
Furthermore, the activation of idle mode, coupled with the discarding of the initial frame following this state, introduces a substantial overhead, thereby constraining the frame rate to half its maximum potential while maintaining the same energy demand.

\begin{figure}[t]
  \centering
  \centerline{\includegraphics[width=8.5cm]{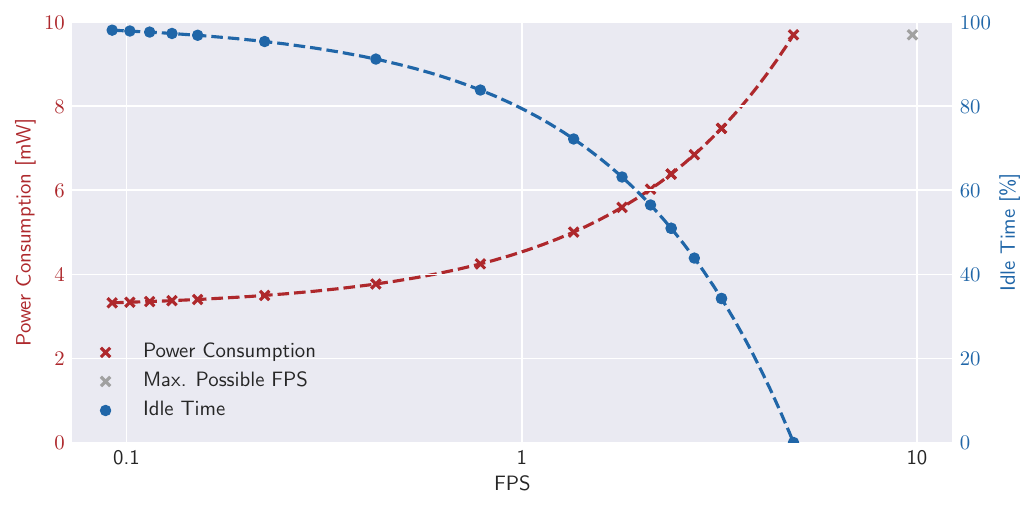}}
\caption{Estimation of the power consumption of the camera module at different frame rates clocked at \SI{5}{\MHz}}
\label{fig:energy}
\end{figure}

A more sophisticated power model of the NanEyeC camera module holds the potential for enhanced power estimation, yet its creation remains particularly challenging. 
The complexity arises from the sensor's multiple settings and variable clock frequencies, which hinders the feasibility of measuring power consumption for all possible configurations. 
The sheer number of settings and frequencies leads to a combinatorial surge in scenarios, rendering exhaustive testing impractical.
The solution to this challenge lies in the development of a model that can generalise accurately from a limited set of measurements.
This necessitates the selection of representative configurations, the utilisation of interpolation or regression techniques, and the consideration of nonlinear interactions between settings.
Moreover, ensuring the reliability of the model across untested scenarios introduces additional complexity, particularly when environmental factors or dynamic workloads influence power consumption.

\section{Conclusion}

This paper presents a \gls{hil} system that facilitates the testing of novel smart camera-based medical device prototypes, such as video capsules, prior to preclinical trials.
The \gls{fpga}-based approach utilizes an image stream from prerecorded medical studies of comparable capsule cameras, employing on-demand fetching to achieve response times of \SI{16.5}{\ms}. 
This capability enables the emulation of the NanEyeC camera module, operating at a clock frequency of \SI{75}{\MHz}. 
Through experimental validation, it is demonstrated that the \gls{hil} ensures the satisfaction of real-time requirements and automatically detects faults during transmission.

The presented system will be used and enhanced for the development of a cutting-edge smart \gls{vce} approach.
To test the robustness of the executed model, image alteration can be utilized by adjusting a case study prior to the simulation.
An extension for altering during the active simulation before injection either within the back-end conversion pipeline or the \gls{fpga} could also be feasible.
By integrating power models of the emulated sensors, it can be possible to roughly estimate energy consumption and, thus, battery life.
Lastly, the \gls{hil} will be later able to receive the capsule's output wireless.
Then, different transmission techniques can be tested as well.


\bibliographystyle{IEEEbib}
\bibliography{paper.bib}

\begin{thebibliography}{10}

\bibitem{joung2013development}
Yeun-Ho Joung,
\newblock ``Development of implantable medical devices: from an engineering
  perspective,''
\newblock {\em International neurourology journal}, vol. 17, no. 3, pp. 98,
  2013.

\bibitem{iddan2000wireless}
Gavriel Iddan, Gavriel Meron, Arkady Glukhovsky, and Paul Swain,
\newblock ``Wireless capsule endoscopy,''
\newblock {\em Nature}, vol. 405, no. 6785, pp. 417--417, 2000.

\bibitem{patel2022obscure}
Apurva Patel, Deepanjali Vedantam, Devyani~S Poman, Lakshya Motwani, and Nailah
  Asif,
\newblock ``Obscure gastrointestinal bleeding and capsule endoscopy: A win-win
  situation or not?,''
\newblock {\em Cureus}, vol. 14, no. 7, 2022.

\bibitem{goran2018capsule}
Loredana Goran, Ana~Maria Negreanu, Ana Stemate, and Lucian Negreanu,
\newblock ``Capsule endoscopy: Current status and role in crohn’s disease,''
\newblock {\em World journal of gastrointestinal endoscopy}, vol. 10, no. 9,
  pp. 184, 2018.

\bibitem{cheung2016usefulness}
Dae~Young Cheung, Jin~Su Kim, Ki-Nam Shim, and Myung-Gyu Choi,
\newblock ``The usefulness of capsule endoscopy for small bowel tumors,''
\newblock {\em Clinical Endoscopy}, vol. 49, no. 1, pp. 21--25, 2016.

\bibitem{chetcuti2021capsule}
Stefania Chetcuti~Zammit and Reena Sidhu,
\newblock ``Capsule endoscopy--recent developments and future directions,''
\newblock {\em Expert review of gastroenterology \& hepatology}, vol. 15, no.
  2, pp. 127--137, 2021.

\bibitem{tziortziotis2021role}
Ioannis Tziortziotis, Faidon-Marios Laskaratos, and Sergio Coda,
\newblock ``Role of artificial intelligence in video capsule endoscopy,''
\newblock {\em Diagnostics}, vol. 11, no. 7, pp. 1192, 2021.

\bibitem{gilabert2022artificial}
Pere Gilabert, Jordi Vitri{\`a}, Pablo Laiz, Carolina Malagelada, Angus Watson,
  Hagen Wenzek, and Santi Segui,
\newblock ``Artificial intelligence to improve polyp detection and screening
  time in colon capsule endoscopy,''
\newblock {\em Frontiers in Medicine}, vol. 9, pp. 1000726, 2022.

\bibitem{werner2023precise}
Julia Werner, Christoph Gerum, Moritz Reiber, J{\"o}rg Nick, and Oliver
  Bringmann,
\newblock ``Precise localization within the gi tract by combining
  classification of cnns and time-series analysis of hmms,''
\newblock in {\em International Workshop on Machine Learning in Medical
  Imaging}. Springer, 2023, pp. 174--183.

\bibitem{farsiu2005multiframe}
Sina Farsiu, Michael Elad, and Peyman Milanfar,
\newblock ``Multiframe demosaicing and super-resolution of color images,''
\newblock {\em IEEE transactions on image processing}, vol. 15, no. 1, pp.
  141--159, 2005.

\bibitem{kaplanMedicalDeviceDevelopment2004}
Aaron~V Kaplan, Donald~S Baim, John~J Smith, David~A Feigal, Michael Simons,
  David Jefferys, Thomas~J Fogarty, Richard~E Kuntz, and Martin~B Leon,
\newblock ``Medical device development: from prototype to regulatory
  approval,''
\newblock {\em Circulation}, vol. 109, no. 25, pp. 3068--3072, 2004.

\bibitem{zwingerCapsoCamSV1PillCam2019}
Lilli~L Zwinger, Britta Siegmund, Andrea Stroux, Andreas Adler, Winfried
  Veltzke-Schlieker, Robert Wentrup, Christian J{\"u}rgensen, Bertram
  Wiedenmann, Felix Wiedbrauck, Stephan Hollerbach, et~al.,
\newblock ``Capsocam sv-1 versus pillcam sb 3 in the detection of obscure
  gastrointestinal bleeding: results of a prospective randomized comparative
  multicenter study,''
\newblock {\em Journal of clinical gastroenterology}, vol. 53, no. 3, pp.
  e101--e106, 2019.

\bibitem{friedrichFirstClinicalTrial2013}
Kilian Friedrich, Sven Gehrke, Wolfgang Stremmel, and Andreas Sieg,
\newblock ``First clinical trial of a newly developed capsule endoscope with
  panoramic side view for small bowel: a pilot study,''
\newblock {\em Journal of gastroenterology and hepatology}, vol. 28, no. 9, pp.
  1496--1501, 2013.

\bibitem{mosleh2020hardware}
Ali Mosleh, Avinash Sharma, Emmanuel Onzon, Fahim Mannan, Nicolas Robidoux, and
  Felix Heide,
\newblock ``Hardware-in-the-loop end-to-end optimization of camera image
  processing pipelines,''
\newblock in {\em Proceedings of the IEEE/CVF Conference on Computer Vision and
  Pattern Recognition}, 2020, pp. 7529--7538.

\bibitem{komorkiewiczFPGAbasedHardwareintheLoopEnvironment2016}
Mateusz Komorkiewicz, Krzysztof Turek, Pawel Skruch, Tomasz Kryjak, and Marek
  Gorgon,
\newblock ``Fpga-based hardware-in-the-loop environment using video injection
  concept for camera-based systems in automotive applications,''
\newblock in {\em 2016 Conference on Design and Architectures for Signal and
  Image Processing (DASIP)}. IEEE, 2016, pp. 183--190.

\bibitem{mascioHardwareLoopImplementation2020}
Chiara~Di Mascio and Giambattista Gruosso,
\newblock ``Hardware in the loop implementation of the oscillator-based heart
  model: a framework for testing medical devices,''
\newblock {\em Electronics}, vol. 9, no. 4, pp. 571, 2020.

\bibitem{RegulationEU20172017}
Publications~Office of~the European~Union,
\newblock ``Regulation (eu) 2017/745 on medical devices,'' Apr 2017.

\bibitem{charoenRhodeIslandGastroenterology2022a}
Amber Charoen, Averill Guo, Panisara Fangsaard, Supakorn Taweechainaruemitr,
  Nuwee Wiwatwattana, Theekapun Charoenpong, and Harlan~G Rich,
\newblock ``Rhode island gastroenterology video capsule endoscopy data set,''
\newblock {\em Scientific Data}, vol. 9, no. 1, pp. 602, 2022.

\bibitem{leflochGalarLargeMultilabel2024}
Maxime Le~Floch, Fabian Wolf, Lucian McIntyre, Christoph Weinert, Albrecht
  Palm, Konrad Volk, Paul Herzog, Sophie~Helene Kirk, Jonas~L Steinhaeuser,
  Catrein Stopp, et~al.,
\newblock ``Galar-a large multi-label video capsule endoscopy dataset,'' 2024.

\bibitem{naneyec}
ams-OSRAM AG,
\newblock {\em NanEyeC Miniture Camera Module}, Oct 2024,
\newblock v5-00.

\bibitem{softwarespi}
Sming,
\newblock {\em Software SPI}, Jan 2022,
\newblock 5.0.0.

\bibitem{zybo2027}
Digilent Inc.,
\newblock {\em ZYBO FPGA Board Reference Manual}, Feb 2017.

\bibitem{ftdi4222}
Future Technology Devices International Ltd.,
\newblock {\em UMFT4222EVUSB2.0 to QuadSPI/I2C Bridge Development Module}, May
  2024,
\newblock 1.5.

\bibitem{PULPFAQs}
Pasquale~Davide Schiavone, Davide Rossi, Antonio Pullini, Alfio Di~Mauro,
  Francesco Conti, and Luca Benini,
\newblock ``Quentin: an ultra-low-power pulpissimo soc in 22nm fdx,''
\newblock in {\em 2018 IEEE SOI-3D-Subthreshold Microelectronics Technology
  Unified Conference (S3S)}, 2018, pp. 1--3.

\bibitem{NexysA7Digilent}
Digilent Inc.,
\newblock {\em Nexys A7 FPGA Board Reference Manual}, Jul 2019.

\end{thebibliography}

\end{document}